\documentclass[
  10pt,                
  letterpaper,         
  aps,prl,             
  reprint,             
  nofootinbib,         
  superscriptaddress,  
  showkeys,            
]{revtex4-1}
\usepackage{graphicx}                                 
\usepackage{xcolor}                                   
\usepackage{array,dcolumn,longtable}                  
\usepackage{amsmath,amssymb,mathtools,slashed}        
\usepackage{textcomp}                                 

\providecommand\lfstyle{}                             
\providecommand\romanup[1]{\text{#1}}                 
\providecommand\greekup[1]{#1}                        

\usepackage[utf8]{inputenc}                           
\usepackage{bm}                                       

\usepackage{siunitx}                                  
\usepackage{hyperref}                                  
\hypersetup{
  hidelinks,  
  breaklinks, 
  unicode     
}
\usepackage[noabbrev]{cleveref}                       

\creflabelformat{equation}{#2#1#3}

\DeclareSIUnit{\fm}{\femto\metre}                     

\newcommand\lhc{\textsc{lhc}}

\newcommand\atlas{\textsc{atlas}}
\newcommand\cms{\textsc{cms}}

\newcommand\dabmod{\textsc{dabm}od}
\newcommand\vusphydro{\text{v-\textsc{usp}hydro}}
\newcommand\ROOT{\textsc{root}}
\newcommand\pythia{\textsc{pythia}}
\newcommand\mckln{\textsc{mckln}}

\newcommand\qcd{\textsc{qcd}}
\newcommand\qgp{\textsc{qgp}}

\newcommand\adscft{\textsc{a}d\textsc{s}/\textsc{cft}}

\renewcommand\vec[1]{\ensuremath{\bm{#1}}}
\newcommand\dimensions[1]{#1\romanup{D}}

\newcommand\piconst{\greekup{\pi}}

\newcommand\dd{\mathop{}\!\romanup{d}}

\DeclarePairedDelimiter\inparen{\lparen}{\rparen}
\DeclarePairedDelimiter\inbrack{\lbrack}{\rbrack}

\newcommand\qcharm{\romanup{c}}

\newcommand\Bmeson{\romanup{B}}
\newcommand\Dmeson{\romanup{D}}

\newcommand\Bzero{\Bmeson^0}
\newcommand\Dzero{\Dmeson^0}

\newcommand\PbPb{\romanup{PbPb}}

\newcommand\snn[1][]{\sqrt{s_\text{NN}}\ifx\\#1\\\else=\SI{#1}{\TeV}\fi}

\newcommand\pt{p_\text{T}}

\newcommand\raa{R_\text{AA}}
\newcommand\vn[1]{v_{#1}}
\newcommand\vnn{\vn{n}}
\newcommand\psin[1]{\psi_{#1}}
\newcommand\psinn{\psin{n}}

\newcommand\Td{T_\text{d}}

\newcommand\tauR{\tau_\text{R}}

\newcommand\fonll{\textsc{fonll}}

\newcommand\cum[1]{\{#1\}}

\newcommand\gammaflow{\varGamma_\text{flow}}
 
\begin{document}
\title{Event-by-event $\vnn$ correlations of soft hadrons and heavy mesons in heavy ion collisions}
\date{\today}

\author{Caio A.~G.~Prado}
\affiliation{Instituto de F\'{i}sica, Universidade de S\~{a}o Paulo, C.P. 66318, 05315-970 S\~{a}o Paulo, SP, Brazil}
\author{Jacquelyn Noronha-Hostler}
\affiliation{Department of Physics, University of Houston, Houston TX 77204, USA}
\affiliation{Department of Physics and Astronomy, Rutgers University, Piscataway, NJ 08854, USA}
\author{Roland Katz}
\affiliation{Instituto de F\'{i}sica, Universidade de S\~{a}o Paulo, C.P. 66318, 05315-970 S\~{a}o Paulo, SP, Brazil}
\author{Alexandre A.~P.~Suaide}
\affiliation{Instituto de F\'{i}sica, Universidade de S\~{a}o Paulo, C.P. 66318, 05315-970 S\~{a}o Paulo, SP, Brazil}
\author{Jorge Noronha}
\affiliation{Instituto de F\'{i}sica, Universidade de S\~{a}o Paulo, C.P. 66318, 05315-970 S\~{a}o Paulo, SP, Brazil}
\affiliation{Department of Physics and Astronomy, Rutgers University, Piscataway, NJ 08854, USA}
\author{Marcelo G.~Munhoz}
\affiliation{Instituto de F\'{i}sica, Universidade de S\~{a}o Paulo, C.P. 66318, 05315-970 S\~{a}o Paulo, SP, Brazil}
\begin{abstract}
  In this paper heavy quark energy loss models are embedded in full event-by-event viscous hydrodynamic simulations to investigate the nuclear suppression factor and the azimuthal anisotropy of $\Dzero$ mesons in $\PbPb$ collisions at $\snn[5.02]$ in the $\pt$ range \num{8}--\SI{40}{\GeV}. In our model calculations, the $\raa$ of $\Dzero$ mesons is consistent with experimental data from the \cms\ experiment.  We present the first calculations of heavy flavor cumulants $\vn2\cum2$ and $\vn3\cum2$ (and also discuss $\vn2\cum4$), which is also consistent with experimental data.  Event-shape engineering techniques are used to compute the event-by-event correlation between the soft hadron $\vnn$ and the heavy meson $\vnn$. We predict a linear correlation between these observables on an event-by-event basis.
\end{abstract}
\maketitle
\section{Introduction} In the last decade, remarkable progress has been made towards understanding the properties of the strong\-ly interacting Quark-Gluon Plasma (\qgp) produced in ultra-relativistic heavy ion collisions~\cite{Gyulassy:2004zy,Shuryak:2004cy}.  A defining feature of the \qgp\ is its ability to flow as a nearly perfect liquid where the shear viscosity to entropy density ratio $\eta/s \sim \num{0.1}$ is an order of magnitude smaller than in ordinary fluids such as water~\cite{Csernai:2006zz}.

Event-by-event relativistic viscous hydrodynamic simulations \cite{Heinz:2013th} have revealed that in this \qcd\ liquid viscous effects are so small that the spatial inhomogeneities present in the initial state are efficiently converted into final state momentum space anisotropy~\cite{Takahashi:2009na,Alver:2010gr}.  One finds that the low-$\pt$ elliptic and triangular flows of all charged particles, $\vn2$ and $\vn3$, are linearly correlated with the corresponding eccentricities of the initial state, $\varepsilon_2$ and $\varepsilon_3$~\cite{Teaney:2010vd,Gardim:2011xv,Niemi:2012aj,Gardim:2014tya}, while higher order flow harmonics exhibit some degree of nonlinear response~\cite{Teaney:2012ke,Niemi:2015qia,Noronha-Hostler:2015dbi,Qian:2016pau}.

In contrast, the physical mechanism responsible for azimuthal anisotropies at high-$\pt$ ($\pt \gtrsim \SI{10}{\GeV}$) relies not on pressure and flow gradients but rather on differences in the path length of highly energetic probes quenched by the expanding medium~\cite{Wang:2000fq,Gyulassy:2000gk}.  This qualitative understanding has been recently confirmed by the first jet energy loss + event-by-event viscous hydrodynamic calculations performed in~\cite{Noronha-Hostler:2016eow,Betz:2016ayq}. A novel feature found in ~\cite{Noronha-Hostler:2016eow,Betz:2016ayq} is that the approximate linear response between $\vn2$ and $\varepsilon_2$ also holds at high $\pt$ on an event-by-event basis.  This implies that the quantum randomness in the position of the nucleons in the incident nuclei, which determines the fluctuations of the initial conditions used in the subsequent hydrodynamic evolution, significantly affects the distribution of path lengths traversed by jets in the medium.

The observation of large azimuthal anisotropy of open heavy flavor mesons~\cite{Abelev:2013lca,Abelev:2014ipa,Adam:2016ssk} adds important new elements to the overall picture discussed above.  Heavy quarks are produced by hard processes in the initial stages of the collision with a non-thermal transverse momentum spectrum, which is expected to relax towards a nearly thermal distribution within a relaxation time scale $\tauR \sim \frac{M}{T} \frac{\eta}{sT}$~\cite{Moore:2004tg} ($M$ is the heavy quark mass).  The difference between the charm and bottom quark masses suggests that, at low-$\pt$, the $\Dzero$ meson $v_n$ should be larger than the corresponding coefficients for $\Bzero$ mesons ~\cite{Nahrgang:2016wig}.  Additionally, at low-$\pt$ quark coalescence~\cite{Greco:2003xt,Greco:2003mm,Fries:2003vb} between heavy and light flavors~\cite{Greco:2003vf,Oh:2009zj} can substantially increase the elliptic flow of heavy mesons~\cite{Nahrgang:2014vza,Cao:2016gvr}, as well as various effects such as Langevin type behavior and hadronic rescattering.

For $\pt \gtrsim \SI{10}{\GeV}$ heavy quarks hadronize mostly via fragmentation\footnote{We neglect possible $\Dmeson$-like states that might form in the \qgp\ \cite{Adil:2006ra}.} and the various low-$\pt$ effects can be neglected.  This provides a simpler scenario for studying how initial state spatial anisotropies are mapped into the final state heavy flavor azimuthal anisotropy.  We address this problem by exploring new techniques that connect soft physics and heavy flavor observables.  A heavy quark energy loss model is embedded on top of event-by-event viscous hydrodynamic backgrounds to study the nuclear suppression factor and $v_n$ of $\Dzero$ mesons in $\PbPb$ collisions at $\snn[5.02]$ for \num{8}--\SI{40}{\GeV}.  We do not intend to find constraints in the chosen energy loss models but rather to discuss the sensitivity of the presented observables that could be used to better understand the physics of heavy quarks on an event-by-event basis.  Our event-by-event approach predicts a linear correlation between the soft hadron and the heavy meson $v_2$ and $v_3$, which could be verified at \lhc.  This linear relationship is not an obvious feature as the anisotropies in the soft and heavy sectors emerge from two very different production, interaction, and hadronization mechanisms. In fact, non-linearities have been already observed in the light sector for higher harmonics, larger viscosities and peripheral collisions~\cite{Betz:2016ayq}.

\section{Details of the model} To simulate the propagation of heavy quark jets in the medium, we developed a modular Monte Carlo code in C++, named \dabmod, combined with \ROOT~\cite{Brun:1997pa} and \pythia8~\cite{Sjostrand:2007gs} libraries, that allows for a variety of energy loss models to be implemented.  Starting from the initial conditions, we sample charm quarks ($\qcharm$) inside the medium with initial momentum distribution given by p\qcd\ \fonll\ calculations~\cite{Cacciari:1998it,Cacciari:2001td} and random initial direction.  We neglect effects of the jets on the medium~\cite{Andrade:2014swa} but the local space-time dependence of the hydrodynamic fields is considered in the energy loss calculations.

For this first study we employ the following simple parametrization for the heavy quark energy loss per unit length \cite{Betz:2014cza} $\frac{\dd E}{\dd x}(T,v) = -f(T,v) \, \gammaflow$, where $T$ is the temperature experienced by the heavy quark, $v$ is the heavy quark velocity, $\gammaflow =\gamma\inbrack[\big]{1-v_\text{flow}\cos(\varphi_\text{quark}-\varphi_\text{flow})}$ with $\gamma = 1\Big/\sqrt{1-v_\text{flow}^2}$, takes into account the boost from the local rest frame of the fluid~\cite{Baier:2006pt}, $\varphi_\text{quark}$ is the angle defined by the propagating jet in the transverse plane, and $\varphi_\text{flow}$ is the flow local azimuthal angle. The jet propagates only in the transverse plane, starting from a production point $\vec{x}_0$, moving in the direction defined by $\varphi_\text{quark}$.

We investigate the cases where $f(T,v)  = \xi T^2$ and $f(T,v) = \alpha$, with $\xi$ and $\alpha$ being constants.  Both forms are simplified approximations for the interactions between heavy quarks and the strongly interacting \qgp . The first choice is inspired by conformal \adscft\ calculations~\cite{Gubser:2006bz} (for non-conformal plasma see \cite{Rougemont:2015wca}).  The second choice is inspired by Ref.\ \cite{Das:2015ana} which showed that a non-decreasing drag coefficient near the phase transition is favored for a simultaneous description of heavy flavor $\raa(\pt)$ and $\vn2(\pt)$ (this is also supported by $T$-matrix calculations~\cite{vanHees:2007me,Riek:2010fk}).  These models are fairly simple to implement and they give a good description of $\raa$.

Our calculations use hydrodynamical profiles to provide the temperature and flow fields at each time step for each event.  We generate hydrodynamic profiles using the $\dimensions{2}+1$ event-by-event relativistic viscous hydrodynamical model, \vusphydro~\cite{Noronha-Hostler:2013gga,Noronha-Hostler:2014dqa,Noronha-Hostler:2015coa}, which passes standard accuracy tests \cite{Marrochio:2013wla}.  Our setup is the same as \cite{Noronha-Hostler:2016eow,Noronha-Hostler:2015coa,Noronha-Hostler:2015coa} (\mckln\ initial conditions at $\PbPb$ $\snn[5.02]$~\cite{Drescher:2006pi,Drescher:2007ax,Drescher:2006ca}, $\eta/s = \num{0.05}$, and initial time $\tau_0 = \SI{0.6}{\fm}$), which describes experimental data in the soft sector, such that all the hydrodynamic parameters are fixed in the present study.  The heavy quarks are evolved on top of hydrodynamic backgrounds until they reach the jet-medium decoupling temperature $\Td$ below which hadronization is performed using the Peterson fragmentation function \cite{Peterson:1982ak}.  This parameter encodes the large uncertainties regarding hadronization of jets in the \qgp\ and is set to vary between $\Td = \SI{120}{\MeV}$ and $\Td = \SI{160}{\MeV}$, inspired by~\cite{Betz:2016ayq,Nahrgang:2016lst,Becattini:2012xb}.  The parameter for the Peterson fragmentation function is fixed so that the heavy meson spectra matches \fonll\ calculations. Coalescence is not taken into account but it will be implemented in future work to extend the validity of our calculations to  low-$\pt$. Hadronic rescattering is not significant at high-$\pt$~\cite{Bratkovskaya:2015foa,Song:2015ykw} and is neglected here.

In this work, an event-by-event analysis is possible by oversampling each individual hydro event with millions of heavy quarks.  From each individual event $\raa^\qcharm(\pt,\varphi)$ for charm quarks is obtained and the corresponding azimuthal coefficients
\begin{equation}\label{eq:vn}
  \vnn^\qcharm(\pt) = \frac
    {\frac{1}{2\piconst} \int_0^{2\piconst}\! \dd\varphi\, \cos\inbrack[\big]{n\varphi-n\psi_n^\qcharm(\pt)} \, \raa^\qcharm(\pt,\varphi)}
    {\raa^\qcharm(\pt)} \,,
\end{equation}
are calculated based on~\cite{Poskanzer:1998yz} with
\begin{equation}\label{eq:psin}
  \psinn^\qcharm(\pt) = \frac{1}{n} \tan^{-1}
    \inparen[\Bigg]{\frac
      {\int_0^{2\piconst}\! \dd\varphi\, \sin(n\varphi)\, \raa^\qcharm(\pt,\varphi)}
      {\int_0^{2\piconst}\! \dd\varphi\, \cos(n\varphi)\, \raa^\qcharm(\pt,\varphi)}
    } \,.
\end{equation}

In reality, there are very few heavy quarks per event and our approach gives the probability for an event with a certain $\vnn$ and $\psinn$ in the soft sector to produce the heavy flavor quantities $\vnn^\qcharm(\pt)$ and $\psinn^\qcharm(\pt)$.  To compare with experimental data the scalar product method~\cite{Luzum:2012da,Luzum:2013yya} is used to calculate the 2 and 4-particle cumulants with the inclusion of multiplicity weighting and centrality class re-binning as described in~\cite{Betz:2016ayq,Gardim:2016nrr}.  We have at least a couple of thousand hydrodynamic events in each centrality class and we checked that our statistical error bars (computed using jackknife resampling \cite{Miller:1974??}) are on the order of $10^{-4}$.

The free parameters $\xi$ and $\alpha$ that define our two energy loss scenarios are determined by matching our model calculations for $\Dzero$ $\raa$ to experimental data at $\pt \gtrsim \SI{10}{\GeV}$ in the 0--10\% centrality class.  The same procedure has been used to fix the jet-medium coupling in the light sector in~\cite{Betz:2014cza,Noronha-Hostler:2016eow,Betz:2016ayq}.  The parameters must be fixed for every decoupling temperature $\Td$ separately.  With these parameters fixed\footnote{We use $\xi = (\num{16.000}, \num{18.500})$ and $\alpha = (\num{0.764}, \num{1.087})\,\si{\MeV}$, where the first term in the parenthesis represents $\Td = \SI{120}{\MeV}$ while the second represents $\Td = \SI{160}{\MeV}$.} we can perform a full simulation across all $\pt$ and centralities.

\begin{figure}
  \centering
  \includegraphics[width=0.4\textwidth]{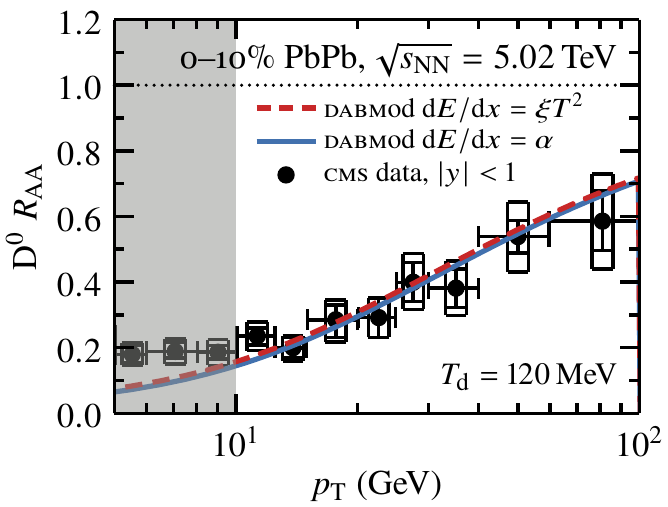}
  \caption{(Color online) $\raa$ for $\Dzero$ mesons in 0--10\% $\PbPb$ collisions at $\snn[5.02]$ and decoupling temperature $\Td = \SI{120}{\MeV}$ comparing two energy loss models with experimental data from \cms\ collaboration~\cite{CMS:2016nrh}.}
  \label{fig:raa}
\end{figure}

\section{Numerical results} Fig.\ \ref{fig:raa} shows our results for $\Dzero$ $\raa$, together with run 2 \lhc\ \cms\ data \cite{CMS:2016nrh}. In this plot we use the decoupling temperature $\Td = \SI{120}{\MeV}$ and compare the two energy loss scenarios.  One can observe that both energy loss models give the same nuclear modification factor in the $\pt$ range considered.  These results are robust with respect to variations in $\Td$, by appropriately fixing $\xi$ and $\alpha$.  Finally, since we do not use the same $\pt$ dependence, our results differ from previous implementations of \adscft-inspired energy loss calculations such as \cite{Akamatsu:2008ge,Horowitz:2007su}.

\begin{figure}[t]
  \centering
  \includegraphics[width=0.4\textwidth]{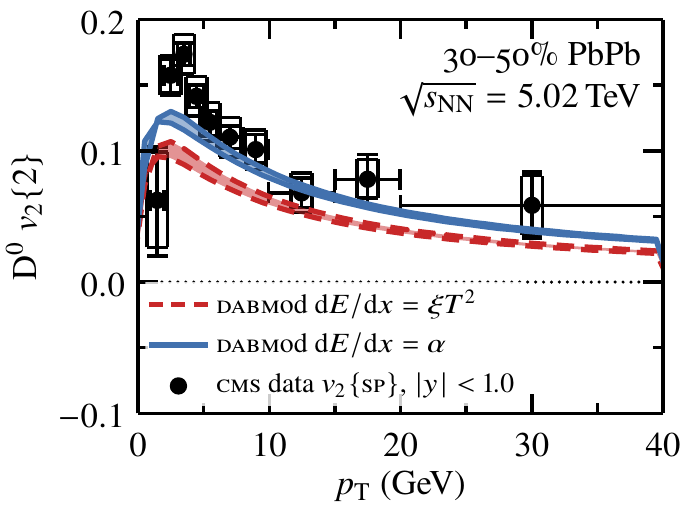}
  \caption{(Color online) $\vn2\cum2$ of $\Dzero$ mesons in 30--50\% $\PbPb$ collisions at $\snn[5.02]$ comparing two energy loss models with \cms\ data \cite{CMS:2016jtu}.  The band corresponds to the decoupling temperature interval $\SI{120}{\MeV} \leq \Td \leq \SI{160}{\MeV}$.}
  \label{fig:cumulantsV2}
\end{figure}

We compute the differential $\vn2\cum2$ \cite{Noronha-Hostler:2016eow,Betz:2016ayq} for $\Dzero$ meson at 30--50\% centrality and compare it to experimental data in Fig.\ \ref{fig:cumulantsV2}.  Our model is consistent with the data at high-$\pt$, though it falls below it at low-$\pt$ where effects such as coalescence, shadowing, and stochastic dynamics could play a significant role.  The model where $\dd E/\dd x \sim \alpha$ gives the largest elliptic flow (as it occurred in~\cite{Das:2015ana}) while the bands illustrate the dependence of these observables with $T_d$. We find that $\vn2\cum2$ increases when $\Td$ is lowered from \SI{160}{\MeV} to \SI{120}{\MeV}, which is expected due to the larger time available to build up the azimuthal anisotropy.  Moreover, we note that even though the chosen $\Td$ range is large, our results are quite robust concerning this parameter.

\begin{figure}[b]
  \centering
  \includegraphics[width=0.4\textwidth]{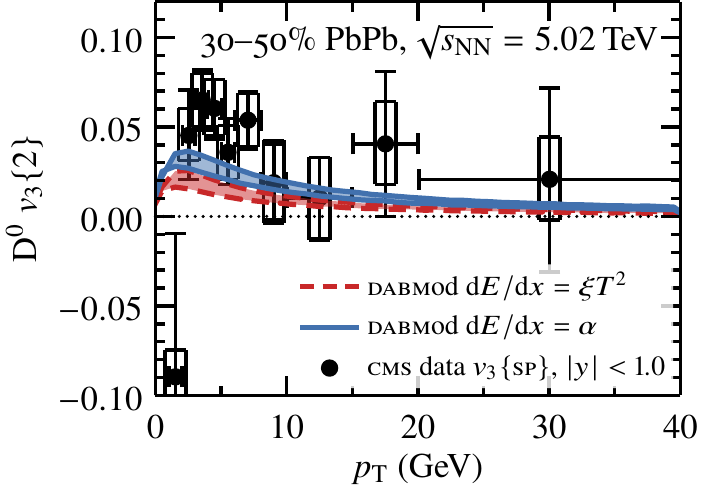}
  \caption{(Color online) $\vn3\cum2$ of $\Dzero$ in 30--50\% $\PbPb$ collisions at $\snn[5.02]$ comparing two energy loss models with \cms\ \cite{CMS:2016jtu}.  The band corresponds to the decoupling temperature interval $\SI{120}{\MeV} \leq \Td \leq \SI{160}{\MeV}$.}
  \label{fig:cumulantsV3}
\end{figure}

Our calculations for the corresponding $\vn3\cum2$ of $\Dzero$ can be found in Fig.\ \ref{fig:cumulantsV3}, together with experimental data for comparison.  $\vn3\cum2$ is a factor $\sim \num{3}$ smaller than the corresponding $\vn2\cum2$ shown in Fig.\ \ref{fig:cumulantsV2} and falls slightly below the experimental data at low-$\pt$.  By considering event-by-event simulations, one sees that a non-decreasing drag coefficient \cite{Das:2015ana} gives not only the largest $\vn2\cum2$ but also the largest $\vn3\cum2$, thereby amplifying the survival of the initial state fluctuations perceived by heavy flavor. Also, $\vn3\cum2$ is particularly sensitive to the decoupling between heavy quarks and the medium in comparison with $\vn2\cum2$ as the bands almost overlap.  We find a slightly smaller $\Dzero$ triangular flow than Refs.~\cite{Nahrgang:2014vza,Nahrgang:2016lst}, which is likely due to the event-plane decorrelation effect present at high $\pt$~\cite{Jia:2012ez,Noronha-Hostler:2016eow,Betz:2016ayq}.  Finally, our calculations in Fig.\ \ref{fig:raa}--\ref{fig:cumulantsV3} show that $\vnn$ are more sensitive to the energy loss models than $\raa$.

Azimuthal anisotropy fluctuations can be systematically investigated using multi-particle cumulants~\cite{Bilandzic:2010jr,Bilandzic:2013kga} which, in the present context of heavy-light flavor flow correlations, should give valuable information about the spectrum of path length fluctuations of heavy quark jets in the medium. However, it is not clear whether current statistics allows for proper measurement of higher order heavy flavor cumulants at high $\pt$ at \lhc\ (or even $\vn3\cum2$ with small enough error bars). Because we oversample each event we have enough statistics to compute multi-particle cumulants such as $\vn2\cum4$ which measures the correlation between a heavy flavor candidate and 3 soft particles \cite{Betz:2016ayq}. For the 30--50\% centrality class and $\pt$ range \num{8}--\SI{40}{\GeV} we find that $\vn2\cum4/\vn2\cum2 \sim \num{0.95}$ within statistical error bars regardless of variations in the energy loss, $\Td$, and quark flavor. A similar value, in the case of light flavor jets, was reported in~\cite{Betz:2016ayq}. The $\vn2\cum4 / \vn2\cum2$ ratio is related to the variance of the $\vn2^2(\pt)$ distribution, which reflects the event-by-event fluctuations in the hard sector due to the initial density fluctuations within our model.  Therefore, this ratio probes the magnitude of the initial density fluctuations and the role they play in the energy loss process.  A more detailed study, involving the centrality dependence of $\vn2\cum4/\vn2\cum2$ and also the calculation of even higher order heavy flavor cumulants will be presented elsewhere.

\begin{figure}[ht]
  \centering
  \includegraphics[width=0.4\textwidth]{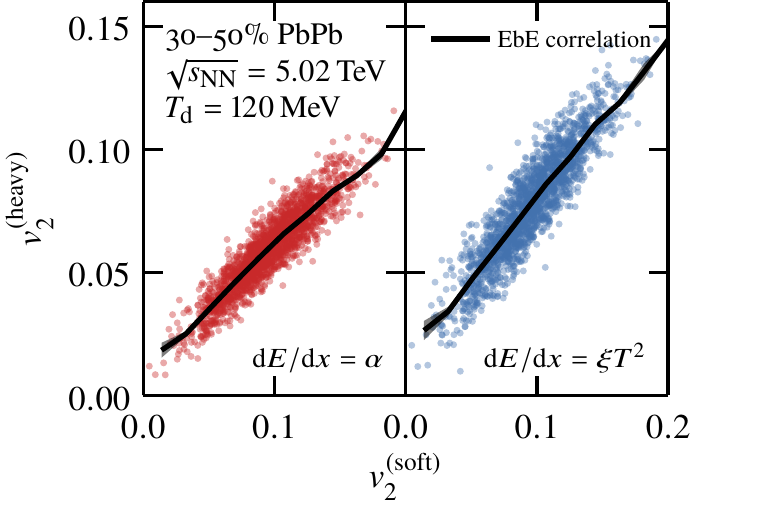}
  \caption{(Color online) Event-by-event correlation between $\Dzero$ $\vn2$ in the $\pt$ range \num{8}--\SI{13}{\GeV} and the soft hadron $\vn2$ for two energy loss models in 30--50\% $\PbPb$ collisions at $\snn[5.02]$.  The solid lines stem from binning the heavy meson result by the soft hadron $\vn2$.}
  \label{fig:distribution}
\end{figure}

We now propose a new observable that encodes the event-by-event fluctuations of heavy flavor $v_n$ at high-$\pt$ that does not require the challenging high statistics needed in multi-particle cumulants analyses.  For a given hydrodynamic event characterized by a soft hadron $\vnn\cum2$, we calculate the corresponding coefficient for $\Dzero$ in each event (this would be akin to the probability that a certain soft event corresponds to this particular heavy flavor $\vnn$), which is shown in the scatter plot in Fig.\ \ref{fig:distribution} for $\vn2$.  This plot shows that the soft and the heavy elliptic flow are correlated event-by-event within a given centrality class. Experimentally, one can bin the soft $\vn2\cum2$ and calculate the corresponding heavy meson $\vn2$ for those set of events, as was done for high $\pt$ identified hadrons by \atlas\ \cite{Aad:2015lwa}. This type of event-shape engineering procedure~\cite{Schukraft:2012ah} gives rise to the solid black lines in Fig.\ \ref{fig:distribution}.  If there were no $\vn2$ fluctuations in the heavy flavor sector one would see a flat, horizontal line. Rather, our calculations predict a clear linear correlation between the heavy meson $\vn2$ and the elliptic flow of all charged hadrons.

\begin{figure}[ht]
  \centering
  \includegraphics[width=0.4\textwidth]{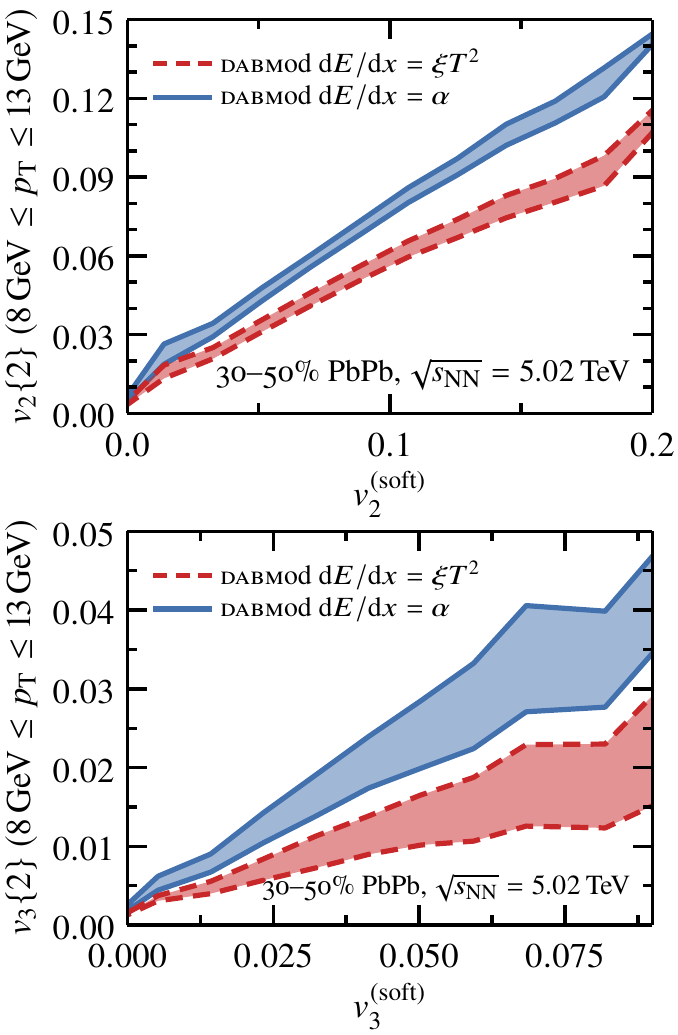}
  \caption{(Color online) Correlation between $\vnn^\text{(heavy)}\cum2$ for $\pt$ in \num{8}--\SI{13}{\GeV} and $\vnn^\text{(soft)}$ for $\Dzero$ in 30--50\% $\PbPb$ collisions at $\snn[5.02]$ using two energy loss models.  The bands represent the variation of the decoupling temperature in $\SI{120}{\MeV} \leq \Td \leq \SI{160}{\MeV}$.}
  \label{fig:correlation}
\end{figure}

This correlation is investigated in detail in Fig.\ \ref{fig:correlation} for $\vn2$ (top panel) and $\vn3$ (bottom panel) where we binned soft vs.~heavy $\vnn$ and varied the energy loss model as well as $T_d$.  Similar separations between the energy loss models are observed for the cumulants, in Figs.\ \ref{fig:cumulantsV2} and \ref{fig:cumulantsV3}, and the correlations, though the latter are highly dependent on the integrated $\pt$ range and should be investigated experimentally in order to determine an optimal choice given the error bars.  Also, even though the widths of the $\Td$ bands are equivalent, $\vn3$ is more sensitive to it than $\vn2$. These results indicate that novel event-shape engineering techniques involving the flow of soft hadrons and heavy flavor will be instrumental in determining the collective behavior of heavy quarks in the \qgp.

\section{Conclusions} In summary, we presented the first theoretical calculations of $\Dzero$ multi-particle flow cumulants in heavy ion collisions performed using realistic event-by-event viscous hydrodynamics.  Our calculations are consistent with the published $\raa$, $\vn2$, and $\vn3$ for $\PbPb$ run 2 \lhc\ data for $\pt=8-40$ GeV.  We computed for the first time the heavy flavor 4-particle cumulant $\vn2\cum4$, which paves the way for experimental and theoretical studies of heavy flavor elliptic flow fluctuations.  

Hydrodynamic viscosity was constrained here by soft bulk flow modeling and further investigation is needed to gauge its effect on our analysis.  In~\cite{Betz:2016ayq} it was observed that an increase from \num{0.05} to \num{0.12} in $\eta/s$ change cumulants by at most 5\%.  Energy loss fluctuations, though not shown in this paper, have been considered following an approach similar to \cite{Betz:2014cza} and we observed that reasonable fluctuations affect the results for the azimuthal coefficients by a few percent.  Further analysis on energy loss fluctuations will be presented elsewhere. 

The linear correlation predicted here between the $\Dzero$ $\vnn$ and the underlying soft hadron $\vnn$ provides a novel signature of collectivity in the heavy flavor sector event-by-event.  Experimental confirmation of this behavior requires extending the current cutting edge event-shape engineering techniques~\cite{Aad:2015lwa,Adam:2015eta} to consider soft-heavy correlations, which should be feasible during the \lhc\ run 2 and 3. This would not only allow for a comparison between the azimuthal anisotropies of heavy quarks and charged hadrons on an event-by-event basis but also give new insight into how the ubiquitous quantum mechanical fluctuations present in the initial state affect the energy loss experienced by heavy quarks in the plasma.

\section*{Acknowledgments} The authors thank Funda\c{c}\~ao de Amparo \`a Pesquisa do Estado de S\~ao Paulo (\textsc{fapesp}) and Conselho Nacional de Desenvolvimento Cient\'ifico e Tecnol\'ogico (\textsc{cnp}q) for support. J.N.H. was supported by the National Science Foundation under grant \textnumero\ \textsc{phy-1513864} and she acknowledges the use of the Maxwell Cluster and the advanced support from the Center of Advanced Computing and Data Systems at the University of Houston to carry out the research presented here. J.N. thanks the University of Houston and Rutgers University for their hospitality.


\bibliography{library}

\end{document}